\documentclass{aa}
\usepackage{mathptm}
\usepackage{psfig}

\begin{document}

   \thesaurus{06 	    
              (08.07.2;     
	       08.13.2;     
               08.16,2;     
	       08.02.3;     
               08.23.1)}    
 
   \headnote{Letter to the Editor} \title{Formation of undermassive
   single white dwarfs and the inf\/luence of planets on late stellar
   evolution}

   \author{G. Nelemans\inst{} \& T. M. Tauris \inst{} }

   \offprints{gijsn@astro.uva.nl}

   \institute{Astronomical Institute, ``Anton Pannekoek'', University
              of Amsterdam, Kruislaan 403, NL-1098 SJ Amsterdam, The
              Netherlands }

   \date{Received \today}

   \maketitle \markboth{G. Nelemans \& T.M. Tauris: Formation of
   undermassive white dwarfs}{}

\begin{abstract} We propose a scenario to form low-mass, single, slow rotating
white dwarfs from a solar-like star accompanied by a massive planet, or a brown
dwarf, in a relatively close orbit (e.g. HD~89707). Such white dwarfs were
recently found by Maxted \& Marsh (1998). When the solar-like star ascends the
giant branch it captures the planet and the subsequent spiral-in phase
expels the envelope of the giant leaving a low-mass helium white
dwarf remnant. In case the planet evaporizes, or fills its own
Roche-lobe, the outcome is a single undermassive white dwarf.
The observed distribution of planetary systems supports the applicability
of this scenario.
 
      \end{abstract}

\section{Introduction}
Recent searches for double degenerates (two white dwarfs in  a binary; Marsh 1995; Marsh, Dhillon \& Duck 1995)
have resulted in the discovery of two single, low-mass helium white dwarfs
-- cf. Table~\ref{props}. Similar undermassive white dwarfs \linebreak[4] ($\la 0.5 M_{\sun}$)
are usually found in binaries and can not be formed from normal, single star
evolution which leaves a $\ga 0.6 M_{\sun}$ \linebreak[4] {C-O} white dwarf as a remnant.
Any potential single, low-mass progenitor star of these newly discovered 
undermassive white dwarfs can be excluded, since they would have a main 
sequence lifetime exceeding the age of our Milky Way.
A scenario has been proposed (Iben, Tutukov \& Yungelson 1997) in which a 
double degenerate has merged, due to the emission of gravitational wave
radiation. According to Maxted \& Marsh (1998), this
scenario predicts high ($\sim 1000 \, {\rm km \;s}^{-1}$) rotational
velocities for the remnant of the merged objects in contradiction with
their measurements of a maximum projected rotational velocity of only
$\sim 50 \, {\rm km \;s}^{-1}$. Therefore the merger scenario seems
questionable -- unless there is an extremely efficient removal of
angular momentum in the merging process, or the inclination angles for
both these systems are extremely small.

In this letter we suggest a different, simple solution to the formation
of these single, low-mass (undermassive) white dwarfs by investigating
the influence of massive planets, or brown dwarfs, in relatively close
orbits around solar-like stars (Sect.~\ref{planets}). A short discussion
of the consequences of our planetary scenario is given in
Sect.~\ref{discussion}.

\begin{table}[t]
\caption{Properties of the two recently discovered undermassive white dwarfs
-- cf.  Marsh, Dhillon \& Duck 1995; Maxted \& Marsh 1998.}
\label{props}
\begin{tabular}{llll}
\hline
Name & Mass ($M_{\sun}$) & $v_{\rm rot} \,\sin i \, ({\rm km \,s^{-1}})$& d (pc)\\
\hline
WD 1353+409 & 0.40 & $<$ 50 & 130 \\ WD 1614+136 & 0.33 & $<$ 50 & 180\\
\hline
\end{tabular}
\end{table}

\section{Planets around solar-like stars}\label{planets}

\subsection{Introduction}
We propose a scenario in which a solar-like star is surrounded  
by a massive planet, or a brown dwarf, in a relatively close orbit.
When the star evolves on the giant branch it will become big enough to 
capture its planet via tidal forces (cf. Rasio et al. 1996; Soker 1996). 
The planet spirals into the envelope of the giant and a so-called common
envelope phase is initiated. The frictional drag on the planet,
arising from its motion through the common envelope, will lead to
loss of its orbital angular momentum (spiral-in) and deposit of
orbital energy in the envelope. The orbital energy is converted into
thermal and kinetic energy of the envelope which is therefore being
ejected.  The result of this common envelope evolution is determined
by the energy balance and the fate of the planet. As a result of
friction, and the large temperature difference between the envelope of
the giant and the equilibrium temperature of the planet, low-mass
planets evaporize due to heating. In case the planet evaporizes
completely, the outcome will be a single star
with a rotating and reduced envelope -- otherwise we end up with a
planet orbiting the naked core of a giant.  The destiny of this white
dwarf-planet system is determined by the orbital separation.

In this letter we first present the expected outcome of a common envelope 
evolution between a giant and a planet; thereafter we look at the important
question of the onset of this evolution.
We will closely follow the treatment of Soker (1996; 1998),
focusing on the cases where (most of) the envelope is lost in a common 
envelope, leaving a undermassive white dwarf.
Research in this field has been carried out to explain elliptical and
bipolar planetary nebulae (Soker 1996) and the morphology of the
Horizontal Branch in clusters (Soker 1998). 

\subsection{The outcome of the common envelope phase}
Below we outline our scenario in somewhat more detail.  By simply
equating the difference in orbital energy to the binding energy of the
envelope of the giant we can compute the ratio of final to initial
separation (Webbink 1984).  Let $\eta _{\rm ce}$ describe the
efficiency of ejecting the envelope, {\em i.e.} of converting orbital
energy into the kinetic energy that provides the outward motion of the
envelope: $ \Delta E_{\rm bind} \equiv \eta _{\rm ce} \,\, \Delta
E_{\rm orb}$ or (using $m_{\rm p} \ll M_{\rm env}$):
\begin{equation}\label{ce}
  a_{\rm f} \simeq \frac{\eta _{\rm ce}\,\lambda}{2} \;
                   \frac{M_{\rm core}\,m_{\rm p}}{M\,M_{\rm env}}\; R_{\rm g} 
      = f \, \left( \frac{\chi}{1-\chi}\right) \,\frac{m_{\rm p}}{M} \; R_{\rm g}
\end{equation}
where $R_{\rm g}$ is the radius of the
giant star at the onset of the spiral-in phase, $\lambda$ is a weighting factor
($<$ 1.0) for the binding energy of the core and envelope of the giant star,
$\chi \equiv M_{\rm core}/M$,
$m_{\rm p}$ is the planetary mass and $M_{\rm core},
M_{\rm env}$ and $a_{\rm f}$ are the mass of the helium
core and hydrogen-rich envelope of the evolved star ($M=M_{\rm
core}+M_{\rm env}$), and the final separation after all the envelope
is expelled,
respectively. In our calculations we chose $\lambda=0.5$ and
$\eta_{\rm ce}=4$ (cf. Tauris 1996; Portegies Zwart \&
Yungelson 1998) and hence $f=1$.

To model the effect of planetary evaporation we follow \linebreak[4] Soker~(1998) and equate
the local sound speed in the giants envelope to the escape velocity
from the (gaseous) planet surface in order to find the approximate location of
evaporation:
\begin{equation}
  c_{\rm s}^2 \approx v_{\rm esc}^2 \Longleftrightarrow
   \gamma\,\frac{k_{\rm B}T}{\mu \,m_{\rm u}} \approx
   \frac{2\,G\,m_{\rm p}}{\alpha \; r_{\rm p}}
\end{equation}
We use a temperature profile for evolved solar-like stars (cf. Fig.~1)
of $T\approx 1.78\times 10^6\, (r/R_{\sun})^{-0.85}${K}, in the
entire interval of $R_{\rm core} < r < R_{\rm g}$, where $R_{\rm core}$
is the radius of the He-core. During the spiral-in the radius of a giant-gas 
planet, $r_{\rm p}$, may expand slightly ($\alpha \,r_{\rm p}$, $\alpha > 1$)
even though only a small amount of mass ($< 0.1 \,m_{\rm p}$) is
believed to be accreted (Hjellming \& Taam 1991).
\begin{figure}[t]
 \psfig{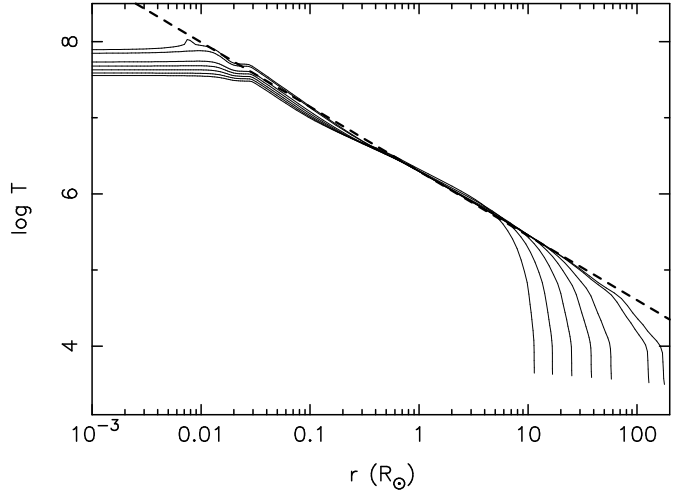}
 \caption{The temperature profiles for 1 $M_{\sun}$ evolved stars on the
          Red Giant Branch. Notice, that log $T$ is approximately a
          linear function of log $r$ at all evolutionary stages from
          the beginning until the tip of the Red Giant Branch (just
          before the helium flash).}
\end{figure}

Solving Eq.~(2), with the temperature dependence given above and
assuming $\gamma = 5/3$ and Pop.I chemical
abundances ({X}=0.7; {Z}=0.02), yields the location of the evaporation:
\begin{equation}\label{a_evap}
  a_{\rm evap} = \left[10\, \alpha\, \left(\frac{M_{\rm J}}{m_{\rm
  p}}\right) \right]^{1.18} \;\; R_{\sun}
\end{equation}
where $M_{\rm J}=0.001 M_{\sun}$ ($\approx$ a Jupiter mass) and we
have assumed $r_{\rm p} = 0.1 R_{\sun}$, which is a reasonable
assumption for all planets and brown dwarfs in the mass range
$0.0001 < m_{\rm p}/M_{\sun} < 0.08$ \linebreak[4] (Hubbard 1994).

For a given stellar structure ({\em i.e.} core and envelope mass and
radius) the final outcome of the common envelope phase is determined
only by the mass of the planet. 
We can easily compute the critical planetary mass for which the planet
evaporizes just at the moment the envelope is completely expelled,
{\em i.e.} when $a_{\rm evap} = a_{\rm f}$. The mass associated with this
critical mass ($m_{\rm crit}$) is found from eqs.~\ref{ce} and \ref{a_evap}
($\alpha=1$):
\begin{equation}\label{mc}
m_{\rm crit} = 10 \; \left[ \left(\frac{1-\chi}{\chi}\right) 
                     \,\left(\frac{M}{M_{\sun}} \right )
                     \,\left(\frac{R_{\rm g}}{100\,R_{\sun}}\right)
                      \right]^{0.46}\;\;M_{\rm J}
\end{equation}
Planets more massive than $m_{\rm crit}$ survive the
spiral-in. However, in order to avoid a destructive mass transfer
to the white dwarf after the spiral-in, it must have a radius smaller
than its Roche-lobe given by (Paczynski 1971):
\begin{equation}
a_{\rm RLO} = \frac{\alpha \,r_{\rm p}}{0.462} \left( \frac{M_{\rm
             WD}}{m_{\rm p}} \right)^{1/3} \, R_{\sun}
\end{equation}
where $M_{\rm WD}=M_{\rm core}$.

If $a_{\rm f} > a_{\rm evap}$ and $a_{\rm f} > a_{\rm RLO}$, the
planet will survive and the entire envelope is lost from the giant
leaving a low-mass helium white dwarf remnant with a planetary
companion. However, if the final separation is small enough, the
planetary orbit will decay due to emission of gravitational
waves on a timescale given by:
\begin{equation}
\tau_{\rm gwr} \approx \frac{ (a_{\rm f}/60\;  R_{\rm WD})^4}
	{(M_{\rm WD}/M_{\sun})^2 \; (m_{\rm p}/M_{\rm J})}
	\; 5.0 \times 10^9 \; \;{\rm yr}
\end{equation}
Hence, also in this case the final outcome of the evolution
might eventually be a single undermassive white dwarf.

Planets less massive than $m_{\rm crit}$ will evaporate (or overflow
their Roche-lobe if $a_{\rm RLO}>a_{\rm evap}$) before the envelope is
expelled completely. However heavy planets deposit significant orbital angular momentum in
the envelope of the giant, causing enhanced mass loss due to
rotation. This could lead to ejection of the envelope by planets
somewhat less massive than $m_{\rm crit}$.  

The change in
structure of the star may alter the further evolution of the giant
considerably. Soker~(1998) suggests that such an evolution could
explain the morphology of the Horizontal Branch in clusters.

For the evolution of the giant we used the relations of Iben \&
Tutukov (1984) for the structure of a (Pop.{I}) giant on the RGB:
$R_{\rm g} = 10^{3.5} M_{\rm core}^4,
\; L = 10^{5.6} M_{\rm core}^{6.5},
\; \dot{M}_{\rm core} = 10^{-5.36} M_{\rm core}^{6.6}$. 
These equations are valid on the RGB for a low-mass star \linebreak[4] ($0.8 \le M/M_{\sun} \le 2.2$).

\subsection{The onset of the common envelope phase: 
            tidal forces and mass loss on the RGB}
The moment the common envelope starts is determined by tidal forces. 
In the absence of any significant tidal interaction the donor star
is only able to capture planets, via Roche-lobe overflow, out to
a distance, $a_{\rm i}^{\rm max} \approx 1.6 \,R_{\rm g}$. 
Taking tidal effects  into account using the equilibrium tide model
(Zahn 1977; Verbunt \& Phinney 1995)
we find, following Soker (1996):
\begin{equation}
  a_{\rm i}^{\rm max} \simeq 2.4 \, R_{\rm g} \, 
                        \left( \frac{1-\chi}{\chi^{9}} \right) ^{1/12} \,
                        \left( \frac{M}{M_{\sun}} \right) ^{-11/12} \,
                        \left( \frac{m_{\rm p}}{10 \, M_{\rm J}} \right) ^{1/8}
\end{equation}
where we have used the equations for the structure of the giant as given above.
In our calculations (see below) we have also included mass loss, which amounts
to as much as $|\Delta M|/M \approx 0.20$ at the tip of the RGB. The mass is
lost as a fast isotropic wind with the specific angular momentum of the giant
causing the orbital separation of the planet to increase by the same ratio
as the total mass of the system decreases. We modeled this effect according to 
the Reimers formula (Kudritzki \& Reimers 1978) with $\eta = 0.6$ 
(cf. Rasio et al. 1996).
 
\subsection{Results}
We will now demonstrate an approximate picture for the fate of stars
with planets of different masses and separations to illustrate the
applicability of this scenario for producing undermassive single white
dwarfs as observed in nature.  For the evolution of a solar-like star
on the Red Giant Branch we will investigate to which separation 
(or alternatively which orbital period) a given planet will be captured by
the star and compute the outcome of the spiral-in process for
different planetary masses.

\begin{figure}[t]
 \psfig{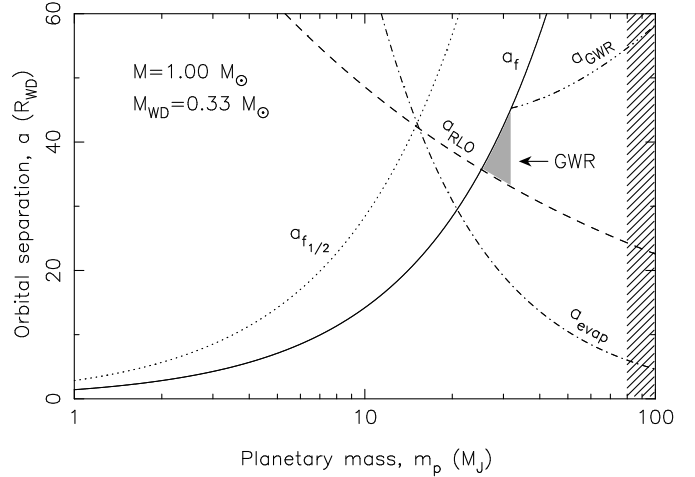}
\caption{Separations of interest (in units of $R_{\rm WD}=10\,000$ km)
after the spiral-in phase for a 1 $M_{\sun}$ star with a core of 0.33
$M_{\sun}$ as a function of planetary mass.
The solid line gives the separation for which the liberated orbital energy
is equal to the binding energy of the envelope (dotted line for ejecting
half of the envelope). 
The dashed line gives the separation below which the planet fills its Roche-lobe.
The dash-dotted line gives the separation at which the planet is
evaporated. A minimum planetary mass of $\sim 21 \,M_{\rm J}$ is needed to 
expel the entire envelope. Planets lighter than this value are seen to be
evaporated. However, for $ 15 < m_{\rm p}/M_{\rm J} < 25$ the planet fills
its Roche-lobe and is likely to be disrupted as a result.
Planets more massive than $\sim 25 \,M_{\rm J}$ survive the common envelope 
phase but will later spiral in due to gravitational wave radiation (shaded
area indicates a spiral-in timescale of less than 5 Gyr). Above 0.08
$M_{\sun}$ (80 $M_{\rm J}$), the companions are heavy enough to ignite
hydrogen as stars (hatched region).}
\label{ma}
\end{figure}

In Fig.~\ref{ma} we have plotted the different critical separations
discussed above as a function of planetary mass. Our example is
based on a 1.0 $M_{\sun}$ star with a core-mass of 0.33 $M_{\sun}$ \linebreak[4] (cf. WD
1614+136 in Table~\ref{props}). We find $m_{\rm crit}=21 \,M_{\rm J}$.
Less massive planets expel only part of the envelope (e.g. a planet
with $m_{\rm p} = 15 \,M_{\rm J}$ will only expel half of the
envelope, neglecting enhanced mass loss of the giant due to the
spin-up of the envelope).
Planets with masses between 15 and $25 \,M_{\rm J}$ are presumably 
disrupted as they fill their Roche-lobe during/after
the spiral-in\footnote{The final fate of the planet depends
on its adiabatic exponent, or actually ($\partial\ln r / \partial\ln m$) and
requires detailed calculations.}. Planets more massive than $\sim 25 \,
M_{\rm J}$ survive the spiral-in and will eject the entire
envelope. However if $m_{\rm p} < 32 \,M_{\rm J}$, the planet will
spiral in, due to emission of gravitational waves, and hence fill its 
Roche-lobe within 5 Gyr.
\begin{figure}
  \psfig{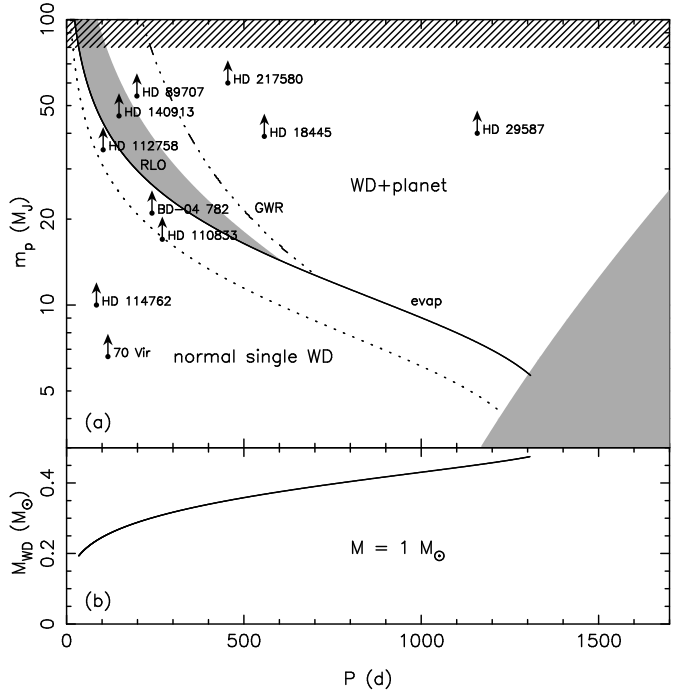}
\caption[]{{\bf a} Final outcome of the common envelope phase for different
planetary masses and initial periods around a 1 $M_{\sun}$ star. The
solid line indicates the critical mass, $m_{\rm crit}$, below which
the planet will evaporize during the spiral-in. Above the solid line
the planet survives the spiral-in phase and the outcome is an
undermassive white dwarf with a planet orbiting it -- unless the
initial period is sufficiently short leading to a disruption of the
planet as it fills its Roche-lobe (left shaded area) after the
spiral-in.  The dash-dotted line indicates the limiting initial
periods below which the planet will fill its Roche-lobe, after the
spiral-in, in less than 5 Gyr due to gravitational wave radiation. 
The dotted line yields the planetary mass for which half of the
envelope is ejected -- neglecting rotation (see text).
Also indicated in the figure are the observed extrasolar planets and brown
dwarfs.  In the shaded area to the right, the planet is too far away
from the giant to be engulfed in its envelope during evolution on the
Red Giant Branch. \newline {\bf b} Final
mass of the white dwarf in case all of the envelope is expelled ({\em
i.e.} $m_{\rm p}>m_{\rm crit}$).}
\label{Pm}
\end{figure}

In Fig.~\ref{Pm} (top) we calculated the final outcome of the
evolution of a planet orbiting a 1 $M_{\sun}$ star as a function of
planetary mass and initial orbital period. We also plotted some of
the known planetary and brown dwarf systems with solar-like stars
(0.70--1.20 $M_{\sun}$). Data were
taken from ``The Extrasolar Planets Encyclopaedia'' ({\small \tt
wwwusr.\-obspm.\-fr/\-departement/\-darc/\-planets/\-encycl.html}).
We notice that, of the observed systems HD~89707 and HD~140913 are the
best candidates for producing single undermassive white dwarfs.
In HD~217580, HD~18445 and HD~29587 the planet is expected to survive
the ejection of the envelope.
In HD~114762 and 70~Vir they are captured already early on the RGB,
where the binding energy of the envelope is too large to be expelled,
so these planets will evaporate shortly after
contact with the evolved donor star. The solitary white dwarfs
resulting from these two systems will therefore be normal {C-O} white dwarfs.\\

\section{Discussion}\label{discussion}
We must bear in mind the uncertainties at work in our scenario, and it is 
possible that future detailed studies of the interactions between a planet
and a common envelope may change the mass limits derived in this letter. 
Also notice that the two undermassive white dwarfs in Table~\ref{props}
might very well have substellar companions (brown dwarfs or planets)
below the observational threshold mass of $\sim 0.1 \,M_{\sun}$.

\subsection{The final mass of the white dwarf}
In the lower panel of Fig.~3 we give the final white dwarf mass in
case all of the envelope is expelled. We see that white dwarfs with
masses between 0.20--0.45 $M_{\sun}$ can in principle be formed with this
scenario. If the common envelope phase initiates while the donor is on
the Asymptotic Giant Branch, a {C-O} white dwarf will be formed.

\subsection{Rotation of the white dwarf}
The final rotational period of the white dwarf is essentially
determined only by the rotation of the core of the giant: the planet
transfers almost all of its angular momentum to the giants envelope 
which is expelled. The rotation of the core strongly
depends on the coupling between the core and the envelope of the giant
(Spruit 1998), but is in any case in agreement with the  the
measured upper-limits for the white dwarfs as given in Table~\ref{props}.

\subsection{A white dwarf ejected from a binary ?}
An other possibility for the formation of single undermassive white
dwarfs is a binary origin. Consider a compact system with a giant star
(the progenitor of the undermassive white dwarf) and a normal white
dwarf companion. When the giant fills its Roche-lobe it transfers its 
envelope to the companion leaving a low-mass helium white dwarf as
a remnant.
The companion may be subsequently lost either because it exploded as
a type Ia SNe, or formed a (high velocity) neutron star from an
accretion induced collapse -- also leading to disruption of the binary.

\acknowledgements{We would like to thank Frank
Verbunt for discussions and Bart Bisscheroux for
providing Fig.~1. This research was supported in part by
the NWO Spinoza-grant SPI 78-327.\\
T.M.T. acknowledges the receipt of a Marie Curie Research Grant
from the European Commission.}

\end{document}